\documentclass{PoS}
\usepackage{amsmath}
\newcommand{\tht}{\textheight}
\newcommand{\ig}{\includegraphics}

\mathchardef\minussign="002D 
\renewcommand{\neg}{\,\minussign\minussign}

\title{Phase Shift with LapH Propagators}

\ShortTitle{stochastic LapH}

\author{J.~Bulava\\
        NIC, DESY, Platanenallee 6, 15738, Zeuthen, Germany\\
        E-mail: \email{jbulava@desy.de}}
\author{J.~Foley\\
        Carnegie Mellon University, Pittsburgh, PA 15213, U.S.A.\\
        E-mail: \email{Justin.Foley@utah.edu}}
\author{\speaker{K.~J.~Juge}\\
        University of the Pacific, Stockton, CA 95211, U.S.A.\\
        E-mail: \email{kjuge@pacific.edu}}
\author{C.~J.~Morningstar\\
        Carnegie Mellon University, Pittsburgh, PA 15213, U.S.A.\\
        E-mail: \email{colin\_morningstar@cmu.edu}}
\author{M.~J.~Peardon\\
        Trinity College, Dublin 2, Ireland\\
        E-mail: \email{mjp@maths.tcd.ie}}
\author{C.~H.~Wong\\
        Carnegie Mellon University, Pittsburgh, PA 15213, U.S.A.\\
        E-mail: \email{chikhimw@cmu.edu}\\\\
\centerline{{\bf For the Hadron Spectrum Collaboration} }}

\abstract{The pion-pion scattering phase shift is computed using LapH propagators. The LapH method for computing quark propagators is used to form two-particle correlation functions with a number of different operators. Excited state energies of two-particle states on 2+1 dynamical, anisotropic lattices ($m_\pi\sim390\ MeV$) are computed to determine the phase shift in the $I=2$ channel. The signal for t-to-t diagrams for the $I=0$ channel are also presented to demonstrate the efficacy of the stochastic LapH method.
}

\FullConference{The XXVIII International Symposium on Lattice Field Theory - Lattice2010\\
June 14 - 19, 2010\\
Villasimius, Italy}

\begin{document}

\section{Introduction}

The determination of the spectrum of excited states in Lattice QCD with light dynamical quarks is hampered by the need to include explicit multiparticle operators in the variational basis of interpolating operators. The need for finite momenta operators seems to require an all-to-all calculation of the quark propagator for the simulation of even the simplest two-pion state. However, it was pointed out \cite{Peardon:2009gh} recently that hadron operators that are constructed from smeared propagators have a natural cutoff in momentum space and that this can be exploited to form one timeslice-to-all quark propagators if the quarks could be smeared. 

This method of computing the quark propagator below the cutoff defined by the quark field smearing (``distillation") was used to compute the $\pi\pi$ scattering length in the $I=2$ channel on dynamical, anisotropic lattices and presented in Williamsburg. We extend this calculation by determining the excited states and the phase shift in this channel using the finite volume method in Euclidean space-time (\cite{Luscher:1986pf}). Furthermore, we modify the distillation algorithm by combining it with a particular stochastic method known as noise partitioning (\cite{Wilcox:1999ab}) or noise dilution. This not only allows a more efficient way to compute all-timeslices-to-all propagators, but it reduces the linear volume dependence of the number of eigenmodes required to achieve the same quark smearing in larger volumes. This was a particular difficulty that had to be solved in the distillation method. 

We present here preliminary results for the $I=2$ phase shift (see \cite{JLQCD} for recent calculations) and the signal for the $t\minussign$to$\minussign t$ pieces of the correlation function in the $I=0$ channel using this new, stochastic smearing algorithm. 

\section{Construction of Operators/Correlators}
\subsection{Quark Field Smearing}
We begin the construction of two-particle correlation functions with the low-mode filtering of the quark fields via the eigenvectors of the 3-dimensional Laplacian operator, $\tilde{\Delta}$. We have computed the low eigenmodes (up to $N_{\rm ev}=128$) of the three-dimensional, gauge covariant Laplacian operator on each timeslice,
$$\tilde{\Delta}v^{(i)}=-\lambda_iv^{(i)}$$
where $\lambda_0$ is the eigenvalue with the smallest magnitude of the lattice Laplacian operator,
$$
\tilde{\Delta}(x,y)=\sum_{k=1}^{3}\left\{\tilde{U}_k(x)\delta(y,x+\hat{k})+\tilde{U}^\dagger_k(x-\hat{k})\delta(y,x-\hat{k})-2\delta(x,y)\right\}.
$$
The {\em tilde} on the gauge fields indicate that they have been smeared using the stout-smearing algorithm (\cite{Morningstar:2003gk}). 

\noindent The quark fields are then smeared using the smearing operator,
$$S_{t_0}(x,x^\prime)=\sum_i\Theta(-\lambda_i+\sigma^2)v^{(i)}(t_0,x)\otimes v^{(i)\dagger}(t_0,x^\prime)$$ and the smeared quark field is given by 
$\tilde{\psi}(t_0,x)=S_{t_0}(x,x^\prime)\psi(t_0,x^\prime).$ 

\noindent We use these quark sources to compute the solution vectors of the Dirac matrix using standard methods. This requires solving for the solutions of each of the eigenvectors used to smear the source field,
$$
\phi^{(i)}(t,x)=M^{\neg1}(t,x;t_0,x_0)v^{(i)}(t_0,x_0).
$$ 
The solution vectors, $\phi^{(i)}(t,x)$, are smeared using the same smearing method but on timeslice $t$,
$$
\tilde{\phi}^{(i)}(t,x)=S_t(x,x^\prime)\phi^{(i)}(t,x^\prime).
$$ 
The smeared quark propagator is then given by,
$$
\tilde{Q}(t,x;t_0,x_0)=S_t(x,x^\prime)M^{\neg1}(t,x^\prime;t_0,x_0^\prime)S_{t_0}(x_0^\prime,x_0).
$$

\section{I=2 Phase Shift}
The quark propagator, $\tilde{Q}(t,x;t_0,x_0)$, is a $t_0$-to-all-$t$ propagator. We can therefore construct pion operators with finite momenta without further inversions (i.e. at no extra cost). We compute the $I=2$ pion scattering phase shift by using several, finite momenta pion operators and diagonalize the correlation matrix to determine the energy eigenvalues of the two-particle state. 
\subsection{Parameters}
We use the $N_f=2+1$ anisotropic lattices with anisotropy tuned to $a_s/a_t=3.5$ (\cite{Edwards:2008ja},\cite{Lin:2008pr}). The lattice spacing in units of $r_0$ is given by  $r_0/a_s=3.221(25)$ and the lattice size was $20^3\times128$. The results are from 90 configurations separated by 40 trajectories. A simple jackknife analysis suggests that the results are independent within the errors. Preliminary results for the two-pion correlation function and scattering length were reported on a smaller lattice, $16^3\times128$ with a pion mass of $\sim390~MeV$ in Ref.~\cite{Bulava:2009ws}. We use the same pion mass and five of the lowest momenta pion operators to project out the S-wave scattering state in the center of mass frame. 

\subsection{$I=2$ Correlation Function}
The $I=2$ correlation function is constructed in the usual way by computing the ``direct" and ``crossed" diagrams. The direct diagram is simply the square of the single pion correlation function,
$$
C_\pi(t,t_0) = \left[\tilde{Q}(t,x;t_0,x_0)\right]^\dagger\tilde{Q}(t,x;t_0,x_0).
$$
\noindent The correlation matrix for the quark exchange diagram is given by

\begin{equation}\nonumber
C^{(cross)}_{ij}(t,t_0)=\left(V_{z^{\prime},t}^\dagger\tilde{M}^{\neg1}_u(z^\prime,t;x^{\prime\!\prime\!\prime},t_0)V_{x^{\prime\!\prime\!\prime}t_0}\right)\left(V_{y^{\prime\!\prime\!\prime}t_0}^\dagger V_{y^{\prime\!\prime\!\prime}t_0}\right)e^{-iq_jy^{\prime\!\prime\!\prime}}\left(V_{x,t}^\dagger\tilde{M}^{\neg1}_u(z^\prime,t;z^{\prime\!\prime\!\prime},t_0)V_{z^{\prime\!\prime\!\prime}t_0}\right)\left(V_{yt}^\dagger V_{yt}\right) e^{ip_iy}\hfill
\end{equation}
\vskip -7mm
\begin{equation}\nonumber \phantom{C^{(cross)}_{ij}(t,t_0)=}\left(V_{z,t}^\dagger\tilde{M}^{\neg1}_u(z,t;x^{\prime\!\prime},t_0)V_{x^{\prime\!\prime}t_0}\right)\left(V_{y^{\prime\!\prime}t_0}^\dagger V_{y^{\prime\!\prime}t_0}\right)e^{iq_jy^{\prime\!\prime}}\left(V_{x^{\prime},t}^\dagger\tilde{M}^{\neg1}_u(x^\prime,t;z^{\prime\!\prime},t_0)V_{z^{\prime\!\prime\!\prime}t_0}\right)\left(V_{y^{\prime}t}^\dagger V_{y^{\prime}t}\right)e^{-ip_iy^{\prime}}\hfill
\end{equation}
where $V_{x,t}$ is the matrix whose columns are the eigenvectors, $v^{(i)}$ and the sum over the momenta is carried out to project out the S-wave. The $I=2$ channel is obtained by subtracting the quark exchange diagram ($C^{(cross)}$) from the square of the single pion correlation function, $C_{\pi,i}C_{\pi,j}$,
$$
C_{i,j}=C_{\pi,i}C_{\pi,j}-C^{(cross)}_{ij}(t,t_0)
$$
The correlation matrix is diagonalized at $t^*/a_t=25$ with the metric timeslice at $t_o/a_t=15$ so as to reduce the contribution from excited states as much as possible without losing the signal to the noise. The stability with respect to $t^*$ and $t_o$ has been checked for the lowest-lying four states. 
\subsection{Extracting the Phase Shift}
\begin{figure}[t]
\begin{center}
\ig[height=0.35\tht,angle=0]{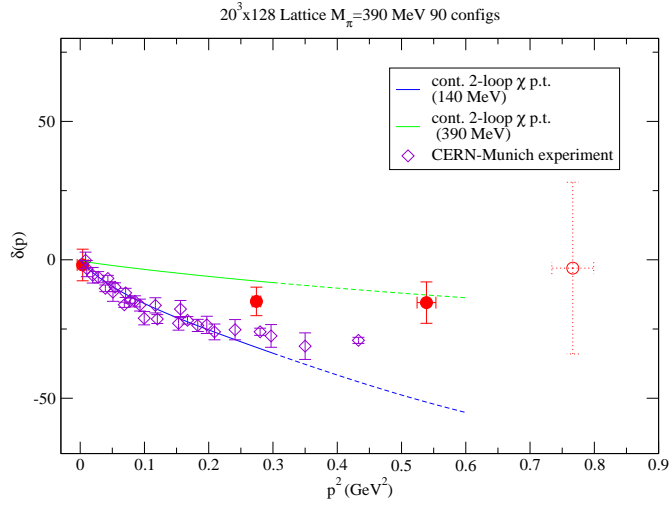} 
\caption{Preliminary results for the $I=2$ phaseshift at $m_\pi=390\ MeV$ (Our points are shown as circles). The two-loop chiral perturbation theory curve was reproduced using the parametrization from Ref.~\cite{Colangelo:2001df} and the $390$ curve was generated by shifting the pion mass.}
\label{fig:phaseshift}
\end{center}
\end{figure}

We follow Luscher's method to compute the phase shift in the infinite volume from the spectra of the two-particle state in a finite volume. First, the physical momenta of the pions is determined from the spectra of the two-particle states and the dispersion relation,
$$
(a_sp_n)^2=\xi^2(a_tm)^2\left[\left(\frac{(a_tE_{\pi\pi})}{2(a_tm)}\right)^2-1\right]
$$
where $\xi=3.5$ is the anisotropy and $a_tm$ is the pion mass (in lattice units) at rest. We then compute the modified Zeta function, $Z_{00}(1;\tilde{n})$ to obtain the phase shift at momentum $a_sp$,
$$
{\rm tan}\delta(p_n)=\frac{\pi^{3/2}\sqrt{\tilde{n}}}{Z_{00}(1;\tilde{n})}\ \ {\rm where}\ \  \tilde{n}=(a_sp_n)^2/\left(\frac{2\pi}{L/a_s}\right)^2.
$$
The results are tabulated in Table~\ref{table:results} and plotted in Fig.~\ref{fig:phaseshift}.
\begin{table}
\begin{center}
\caption{Phase shifts from the two-pion energies.}
\label{table:results}
\begin{tabular}{ccccc}
\hline
 $n$ & $a_tE_{\pi\pi}^{(i)}$ & $a_sp^{(i)}$ & $\tilde{n}$ & $\delta(p^{(i)})$ \\ \hline\hline
0 & 0.150(3) & 0.04(4) & 0.015(33) & -1.8(57) \\\hline
1 & 0.275(4) & 0.326(4) & 1.078(26) & -15(5) \\\hline
2 & 0.368(4) & 0.457(6) & 2.118(58) & -16(7) \\\hline
3 & 0.447(9) & 0.54(1) & 3.01(13) & -3(30) \\\hline
\end{tabular}
\end{center}
\end{table}

\section{$t\minussign$to$\minussign t$ Diagrams}
One of the major challenges in lattice hadron spectroscopy is the evaluation of contributions from disconnected diagrams and box diagrams that appear in correlation functions. These diagrams require the quark propagator from a timeslice $t$, back to $t$ ($t\minussign$to$\minussign t$ diagrams) on some number of timeslices. One way to compute these contributions would be to compute wall propagators from every timeslice $t$ which would require at least $N_t$ times more inversions than before (on our lattices, this factor is roughly 128). While this is not impossible to do in practice, we note that it may be unnecessary as the measurements from neighbouring timeslices may be strongly correlated, and also because the signal diminishes exponentially with time. The other more important reason to combine the LapH method with a stochastic algorithm is to control the number of eigenvectors of the Laplacian that needs to be computed as one progresses to larger and larger lattices. 

\subsection{Stochastic Estimation}
We choose $Z_4$ noise for our stochastic sources which fits naturally with having complex fields on our lattice. It has also been shown in some cases to have smaller variance than other noise choices (\cite{Dong:1993pk}). 
Independent $Z_4$ noise sources, $\varrho_{[i]}$, need to be chosen for each quark line in the correlation function of interest. These have the property
\begin{equation*}
\langle\!\!\langle\varrho\rangle\!\!\rangle=0\ \ {\rm and} \ \ 
\langle\!\!\langle\varrho_{[i]}\varrho^\dagger_{[j]}\rangle\!\!\rangle=\delta_{ij}
\end{equation*}
where the double bracket indicates an average over the noise sources. 
The quark propagator on a given configuration can then be written as,
\begin{eqnarray}\nonumber
Q&=&D_jS_tM^{\neg1}\langle\!\!\langle\varrho\varrho^\dagger\rangle\!\!\rangle S_{t_0}D_k^\dagger\\\nonumber
&=&\langle\!\!\langle D_jS_{t}M^{\neg1}\varrho\,(D_kS_{t_0}\varrho)^\dagger \rangle\!\!\rangle
\end{eqnarray}
where $D_j$ is the covariant displacement operator in direction $\hat{e}_j$.

In order to avoid contaminating the eigenvectors of the Laplacian with our random noise sources, we place the $Z_4$ noise only in the LapH subspace. The noise vectors therefore only have spin and eigenvector indices on each timeslice. The noise vectors will be fully diluted in the time direction for connected diagrams, but can be interlaced in time for the disconnected diagrams. \\
The full dilution scheme is given by
$$
\varrho_{[A]si}(t)\varrho_{[B]s^\prime i^\prime}^\dagger(t^\prime)=\delta_{AB}\delta_{tt^\prime}\delta_{ss^\prime}\delta_{ii^\prime}
$$
(without an average over the noise sources). However, this scheme is very expensive and is unnecessary in practice. The scheme dependence of some observables have been reported elsewhere in these proceedings (\cite{Foley:2010vv}). We only note here that the full time-dilution is usually unnecessary for computing disconnected diagrams since the noise on different timeslices do not interfere with each other as long as they have a separation that is five (a conservative estimate) or greater. This leads to an interlaced dilution scheme which have noise sources on more than one timeslice, but separated by several timeslices. The same can be done for the eigenvector indices and the spin degrees of freedom. 
\subsection{A box diagram example}
We have computed the box diagram that appear in the $I=0$ channel to test the efficacy of the ``stochastic LapH" method. The box diagram can be written in a compact form by using the noise source vectors, $\tilde{\rho}_{[A]s,i}(t_0,\vec{x}_0)$, and the corresponding solutions, $\tilde{\varphi}_{[A]s,i}(t,\vec{x})=M^{\neg1}\tilde{\rho}_{[A]s,i}(t_0,\vec{x}_0)$ as
\begin{equation}
\left(\tilde{\varrho}^{(a)\dagger}_{[0]t}(\vec{x}_1)\gamma_5\tilde{\varphi}^{(d)}_{[3]t}(\vec{x}_1)\right)
\left(\tilde{\varrho}^{(d)\dagger}_{[3]t}(\vec{x}_1^\prime)\gamma_5\tilde{\varphi}^{(c)}_{[2]t}(\vec{x}_1^\prime)\right)
\left(\tilde{\varrho}^{(c)\dagger}_{[2]t_0}(\vec{x}_0^\prime)\gamma_5\tilde{\varphi}^{(b)}_{[1]t_0}(x_0^\prime)\right)
\left(\tilde{\varrho}^{(b)\dagger}_{[1]t_0}(\vec{x}_0)\gamma_5\tilde{\varphi}^{(a)}_{[0]t_0}(\vec{x}_0)\right)
\end{equation}
where the colour indices are contracted within the round brackets and the dilution indices have been combined into one superscript. (This example is given for the usual Dirac matrix.)

\begin{figure}[t]
\begin{center}
\ig[height=0.35\tht,angle=0]{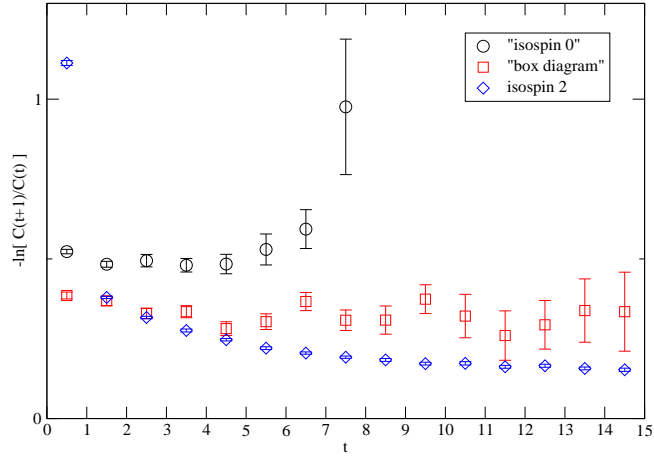} 
\caption{Logarithmic ratios of correlation functions. The data with the label ``I=0" has all contributions to the correlator except for the disconnected diagrams.}
\label{fig:ratios}
\end{center}
\end{figure}

The logarithmic ratios for the $I=0$ correlation function with the box diagram but without the disconnected diagrams is shown in Fig.~\ref{fig:ratios}. We obtain a good signal for the box diagram with with a single timeslice source for the propagator from $t_0$ to $t$ (and interlace 6 for the eigenvectors) and time (interlace 12) and eigenvector (interlace 4) dilutions with full spin dilution. The ratio of correlators of neighbouring timeslices for the $I=2$ and the $I=0$ correlator (without the completely disconnected diagrams) are shown are shown in Fig.~\ref{fig:ratios}. 

\section{Summary}
The LapH quark smearing algorithm (a particular choice of distillation) has been tested for two of the simplest two-particle states. For the $I=2$ channel, the phase shift has been computed with several pion operators with non-zero momenta through L\"uscher's finite volume method. We have obtained a good signal for the first three, lowest lying momenta states in a $(\sim2.5\ {\rm fm})^3$ volume for $m_\pi\simeq390\ MeV$ with $90$ configurations. It is clear that we require lighter pions and larger boxes in order to compare with chiral perturbation theory calculations. These simulations are under way. 

The box diagram in the $I=0$ channel has been computed using a new stochastic algorithm (stochastic LapH) in order to handle the $t\minussign$to$\minussign t$ diagrams. Apart from the fact that this method can extend the LapH algorithm to large lattice volumes, it allows the simulation of $t\minussign$to$\minussign t$ diagrams in an efficient way. We are currently working on the full $I=0$ calculation with disconnected diagrams and the $I=1$ channel to study the $\rho$ decay. 

\section*{Acknowledgements}
This work has been partially supported by National Science Foundation awards PHY-0970137, PHY-0510020, PHY-0653315, PHY-0704171 and through TeraGrid resources provided by Athena at the National Institute for Computational Sciences (NICS) under grant number TG-PHY100027 and NICS and the Texas Advanced Computing Center under TG-MCA075017. M.P. is supported by Science Foundation Ireland under research grant 07/RFP/PHYF168. We thank our colleagues within the Hadron Spectrum Collaboration. These calculations were performed using the Chroma software suite \cite{Edwards:2004sx}.

\end{document}